\documentclass[twocolumn]{aastex631}

\begin{document}

\title{Improved limits on the 21cm signal at $z=6.5-7.0$ with the MWA using Gaussian information}
%\title{Data Distributions for Improved EoR Science}

%\correspondingauthor{August Muench}
%\email{greg.schwarz@aas.org, gus.muench@aas.org}

\author[0000-0001-6324-1766]{Cathryn M. Trott}
\affiliation{International Centre for Radio Astronomy Research, Curtin University, Bentley, WA, Australia}
\affiliation{ARC Centre of Excellence for All Sky Astrophysics in 3 Dimensions (ASTRO
3D), Bentley, Australia}

\author{C.~D.~Nunhokee}
\affiliation{International Centre for Radio Astronomy Research, Curtin University, Bentley, WA, Australia}
\affiliation{ARC Centre of Excellence for All Sky Astrophysics in 3 Dimensions (ASTRO
3D), Bentley, Australia}

\author{D.~Null}
\affiliation{International Centre for Radio Astronomy Research, Curtin University, Bentley, WA, Australia}
\affiliation{Australian SKA Regional Centre (AusSRC), Curtin University, Bentley, WA, Australia}
\affiliation{ARC Centre of Excellence for All Sky Astrophysics in 3 Dimensions (ASTRO 3D), Bentley, Australia} 

\author{N.~Barry}
\affiliation{ARC Centre of Excellence for All Sky Astrophysics in 3 Dimensions (ASTRO 3D), Bentley, Australia} 
\affiliation{School of Physics, The University of New South Wales, Australia}

\author{Y.~Qin}
\affiliation{RSAA, The Australian National University, Mt Stromlo, Cotter Rd, Australia}
\affiliation{ARC Centre of Excellence for All Sky Astrophysics in 3 Dimensions (ASTRO 3D), Bentley, Australia} 

\author{R.~B.~Wayth}
\affiliation{International Centre for Radio Astronomy Research, Curtin University, Bentley, WA, Australia}
\affiliation{ARC Centre of Excellence for All Sky Astrophysics in 3 Dimensions (ASTRO 3D), Bentley, Australia} 
\affiliation{SKA Observatory, SKA-Low Science Operations Centre, 26 Dick Perry Avenue, Kensington WA 6151, Australia}

\author{J.~L.~B.~Line}
\affiliation{International Centre for Radio Astronomy Research, Curtin University, Bentley, WA, Australia}
\affiliation{ARC Centre of Excellence for All Sky Astrophysics in 3 Dimensions (ASTRO 3D), Bentley, Australia} 

\author{C.~H.~Jordan}
\affiliation{International Centre for Radio Astronomy Research, Curtin University, Bentley, WA, Australia}
\affiliation{ARC Centre of Excellence for All Sky Astrophysics in 3 Dimensions (ASTRO 3D), Bentley, Australia} 

\author{B.~Pindor}
\affiliation{School of Physics, The University of Melbourne, Parkville Vic 3010, Australia} 
\affiliation{ARC Centre of Excellence for All Sky Astrophysics in 3 Dimensions (ASTRO 3D), Bentley, Australia} 

\author{J.~H.~Cook}
\affiliation{International Centre for Radio Astronomy Research, Curtin University, Bentley, WA, Australia}
\affiliation{ARC Centre of Excellence for All Sky Astrophysics in 3 Dimensions (ASTRO 3D), Bentley, Australia} 

\author{J.~Bowman}
\affiliation{School of Earth and Space Exploration, Arizona State University, AZ, USA}
%\affiliation{} 
%\affiliation{}

\author{A.~Chokshi}
\affiliation{School of Physics, The University of Melbourne, Australia}
\affiliation{ARC Centre of Excellence for All Sky Astrophysics in 3 Dimensions (ASTRO 3D), Bentley, Australia}

\author{J.~Ducharme}
\affiliation{Department of Physics, Brown University, Providence RI, USA}

\author{K.~Elder}
\affiliation{School of Earth and Space Exploration, Arizona State University, AZ, USA}
%\affiliation{} 
%\affiliation{}

\author{Q.~Guo}
\affiliation{Shanghai Astronomical Observatory, Chinese Academy of Sciences, 80 Nandan Road, Shanghai 200030, China}

\author{B.~J.~Hazelton}
\affiliation{University of Washington, WA, USA}

\author{W.~Hidayat}
\affiliation{Kumamoto University, Japan}

\author{T.~Ito}
\affiliation{Kumamoto University, Japan}

\author{D.~Jacobs}
\affiliation{School of Earth and Space Exploration, Arizona State University, AZ, USA}
%\affiliation{} 
%\affiliation{}

\author{E.~Jong}
\affiliation{International Centre for Radio Astronomy Research, Curtin University, Bentley, WA, Australia}
\affiliation{ARC Centre of Excellence for All Sky Astrophysics in 3 Dimensions (ASTRO 3D), Bentley, Australia} 

\author{M.~Kolopanis}
\affiliation{School of Earth and Space Exploration, Arizona State University, AZ, USA}
%\affiliation{} 
%\affiliation{}

\author{T.~Kunicki}
\affiliation{Department of Physics, Brown University, Providence RI, USA}

\author{E.~Lilleskov}
\affiliation{Department of Physics, University of Washington, WA, USA}

\author{M.~F.~Morales}
\affiliation{Department of Physics, University of Washington, WA, USA}

\author{J.~Pober}
\affiliation{Department of Physics, Brown University, Providence RI, USA}

\author{A.~Selvaraj}
\affiliation{International Centre for Radio Astronomy Research, Curtin University, Bentley, WA, Australia}
\affiliation{ARC Centre of Excellence for All Sky Astrophysics in 3 Dimensions (ASTRO 3D), Bentley, Australia} 

\author{R.~Shi}
\affiliation{Department of Physics, Brown University, Providence RI, USA}
\affiliation{Astronomy Department, The University of Florida, FL, USA} 

\author{K.~Takahashi}
\affiliation{Kumamoto University, Japan}

\author{S.~J.~Tingay}
\affiliation{International Centre for Radio Astronomy Research, Curtin University, Bentley, WA, Australia}

\author{R.~L.~Webster}
\affiliation{School of Physics, The University of Melbourne, Australia}
\affiliation{ARC Centre of Excellence for All Sky Astrophysics in 3 Dimensions (ASTRO 3D), Bentley, Australia} 

\author{S.~Yoshiura}
\affiliation{Mizusawa VLBI Observatory, National Astronomical Observatory of Japan, 2-21-1 Osawa, Mitaka, Tokyo 181-8588, Japan}

\author{Q.~Zheng}
\affiliation{Shanghai Astronomical Observatory, Chinese Academy of Sciences, 80 Nandan Road, Shanghai 200030, China}

\begin{abstract}
We explore the properties of interferometric data from high-redshift 21~cm measurements using the Murchison Widefield Array. These data contain redshifted 21~cm signal, contamination from continuum foreground sources, and radiometric noise. The 21~cm signal from the Epoch of Reionization is expected to be highly-Gaussian, which motivates the use of the power spectrum as an effective statistical tool for extracting astrophysical information. We find that foreground contamination introduces non-Gaussianity into the distribution of measurements, and then use this information to separate Gaussian from non-Gaussian signal. We present improved upper limits on the 21cm EoR power spectrum from the MWA using a Gaussian component of the data, based on the existing analysis from \citet{nunhokee2024}. This is extracted as the best-fitting Gaussian to the measured data. Our best 2$\sigma$ (thermal+sample variance) limit for 268 hours of data improves from (30.2~mK)$^2$ to (23.0~mK)$^2$ at $z=6.5$ for the EW polarisation, and from (39.2~mK)$^2$ to (21.7~mK)$^2$ = 470~mK$^2$ in NS. The best limits at $z=6.8$ ($z=7.0$) improve to $P < (25.9~$mK)$^2$ ($P < (32.0$~mK)$^2$), and $k = 0.18h$/Mpc ($k = 0.21h$/Mpc). Results are compared with realistic simulations, which indicate that leakage from foreground contamination is a source of the non-Gaussian behaviour.
\end{abstract}



\section{Introduction} \label{sec:intro}
The first billion years of the Universe witnessed the birth of the first stars and galaxies, and the tranformation of the dominant hydrogen component of the neutral intergalactic medium, to ionized. Data from the James Webb Space Telescope (JWST) over the past two years have opened a new observational window on this period, probing details of the first generations of galaxies on galactic and cluster scales. In that time, we have learned that massive galaxies existed earlier than previously understood \citep{tang21,harikane23,fujimoto23,boyett24}, and that there appears to be an overdensity of detected Lyman-$\alpha$ emitting galaxies relative to the expected distribution of neutral hydrogen in the IGM \citep{tang24}. Measurements of the ionizing efficiency of early galaxies, and hence the fraction of ionizing photons that escape them to ionize the IGM, also exceed measurements from other tracers, leading to a tension in the observations of the early parts of Reionization \citep{munoz24}, and pointing to a model where Reionization started early. These optical/IR observations motivate improved measurements of the IGM neutral hydrogen itself, via its 21cm hyperfine emission line, as the tracer that provides the radiative and thermal conditions of the IGM as it responds to the ionizing light from the galaxies. While the JWST (and other) observations provide the small-scale information about galaxies, and limited information about the IGM through Ly-$\alpha$ measurements, 21cm studies with low-frequency radio telescopes provide the large-scale information directly about the IGM. Progress in both of these observational fields therefore provides the complete and complementary picture of the early evolution of the Universe.

Several international experiments have been pursuing this redshifted IGM signal using low-frequency radio interferometers, noting that the rest 21cm emission is redshifted to 1.5--2.0~m (50--200~MHz). These include the Murchison Widefield Array, MWA{\footnote[1]{http://www.mwatelescope.org}} \citep{bowman13_mwascience,tingay13_mwasystem,jacobs16}; the Precision Array for Probing the Epoch of Reionization, PAPER{\footnote[2]{http://eor.berkeley.edu}} \citep{parsons10}; the LOw Frequency ARray, LOFAR{\footnote[3]{http://www.lofar.org}} \citep{vanhaarlem13,patil16}; the Long Wavelength Array, LWA{\footnote[4]{http://lwa.unm.edu}} \citep{ellingson09}, HERA \citep{deboer16,berkhout24}, and the under construction SKA-Low \citep{koopmans15}. These experiments all have unique challenges, and approach the observations and analysis in different ways. They are also optimised to produce their best results at different redshifts and wavenumbers. As such, together they provide more independent constraints on the early Universe. Results from all three current experiments have ruled out certain cold Reionization models, whereas HERA results at $z=10.4$ have provided the most stringent constraints on the heating of the IGM above the adiabatic limit at early times \citep{hera23}. 

The brightness temperature fluctuations of the redshifted 21cm line as a function of scale are expected to be highly Gaussian-distributed early in Cosmic Dawn before strong non-linear evolution and small-amplitude higher-order terms emerge later in the Reionization epoch \citep{wyithe07}, although note that this work made over-simplifying assumptions about the signal (e.g., x-ray heating effects). The variance (second-order moment) therefore contains \textit{most} of the information across time, thereby motivating use of the power spectrum as a useful statistic to probe this epoch, and provide astrophysical constraints on conditions at early times. Even at late times, non-Gaussianity of the 21~cm is difficult to detect, with only the SKA projected to be able to explore it at high signal-to-noise ratio \citep{cook24,ma23}. \citet{liu20} reviewed a range of approaches to the analysis of 21cm interferometric datasets, discussing the properties that can be used. More recently, \citet{fronenburg24} explored non-Gaussianity of foregrounds in their work to use cross-correlations to extract 21cm signal.

The spherically-averaged power spectrum encodes the distribution of signal power as a function of spatial scale. Several recent and current experiments have reported upper limits on the 21cm cosmological power at different redshifts, with none yet claiming a detection. From the LOFAR telescope, with its core based in the Netherlands, the most recently reported result at $z=9.1$ improved on an earlier analysis of the same dataset with an upper limit of $\Delta^2_{21} <$ (80 mK)$^2$ at $k = 0.075$~h/Mpc \citep{acharya24}. From the HERA telescope, based in South Africa, the most recently reported reported results were ($z=7.9$) $\Delta^2_{21} < 457$ mK$^2$ at $k = 0.34$~h/Mpc, and ($z=10.4$) $\Delta^2_{21} < 3,496$ mK$^2$ at $k = 0.36$~h/Mpc \citep{hera23}. In \citet{trott20} 110 hours of MWA data were used to place a limit of $\Delta^2_{21} < $(45 mK)$^2$ at $k = 0.075$~h/Mpc at $z=6.5$. This was then improved in \citet{nunhokee2024}, where a set of 268 hours of data were used to report a 2$\sigma$ upper limit of $\Delta^2_{21} <$ (30.2 mK)$^2$ at $k = 0.17$~h/Mpc at the same redshift. These data are used in this work.

%Despite the spherically-averaged (1D) power spectrum providing the largest accumulation of data for the isotropically-distributed signal, often the two-dimensional (2D) power spectrum is used as an intermediate product, whereby the data are retained in angular ($k_\bot$) and line-of-sight ($k_\parallel$) modes. This is to attempt to separate the line emission signal from continuum foregrounds, for which the contamination should be contained in the low $k_\parallel$ modes. The point source population is expected \citep{datta10,trott12,vedantham12,liu15,murray17} and observed \citep{jacobs16,trott16,ali15,beardsley16} to form a wedge-like feature in the low $k_\parallel$ modes, due to the incomplete sampling of a radio interferometer. Despite this `foreground avoidance' technique avoiding >99\% of the foreground emission, chromatic instruments, imprecise calibration, incomplete source models, and limited bandwidth, all combine to allow leakage further into the higher $k_\parallel$ modes, biasing the signal and currently limiting a detection.

In \citet{kde2019} we used a Kernel Density Estimator (KDE) to extract information about the distribution of visibility data from 21cm observations. We explored the properties of the data across real and imaginary components of visibilities, and also other measures of the difference between distributions. In this work, we apply such techniques to a large dataset of 268 hours of data from the EoR0 observing field \citep{nunhokee2024}, showing that we can obtain improved upper limits on the power spectrum at $z=6.5-7.0$. We study the distributions of data with a view to extracting second-order statistics of the cosmologically-redshifted 21cm signal. Despite growing non-Gaussianity near the end of Reionization, we expect the 21cm signal to be highly-Gaussian in $k$-space, where each measurement provides a realisation of the 21cm field at a particular spatial scale (i.e., for a given spatial scale, the individual measurements in 3D $k$-space can be used to form an empirical distribution, which we expect to be zero-mean and Gaussian). Non-gaussianity of the 21cm signal at $z<7$ is not expected to be detectable with less than 1000 hours of MWA observations \citep{ma23}. We use the distributions of the data to extract the Gaussian component of the measured signal, potentially avoiding some of the foreground contaminants that may have non-Gaussian components.

Exploration of the distributions of datasets can be undertaken at different stages of processing. At the basic level, each calibrated visibility contributing to a particular spatial mode can be used to estimate the underlying distribution via some standard technique; here Kernel Density Estimators are employed. While this retains maximal information, it comes at the cost of coherent integration, because each measurement is treated as an independent measure; effectively producing a fully-incoherent power spectrum estimate. An intermediate approach is to estimate the distributions from the coherently-averaged data; i.e., take the gridded visibilities and use those averaged values as the realisations from which to estimate the underlying distribution. This approach is justified if, (1) there are sufficient independent samples (i.e, uv-cells with data) for a robust estimate; (2) there is a good distribution of angles sampled (i.e., a well-filled uv-plane) for the ensemble of measurements to reflect the full distribution. Both of these requirements are justified for large integrations of data with the MWA telescope, which has excellent instantaneous and synthesised uv-coverage.

There are \textit{a priori} properties of the Gaussian fits that would provide confidence that the results are isolating the 21cm signal (and residual Gaussian foregrounds) while removing the bulk of the contamination: (1) consistent power between the real and imaginary components of the visibilities; (2) consistent power between the NS and EW polarisations at a given scale; (3) consistent power between two independent observing fields in the sky, for a comparable amount of data. The first two can be verified with the data used in this work, whereas the latter will need to wait for deep analysis of a separate observing field. Residual Gaussian-distributed foreground contamination is expected to remain in the data, and therefore this approach will only be partly successful in removing non-21cm signal components. There is no unique decomposition of data into a Gaussian and non-Gaussian component, or multiple Gaussian components, due to the convolution properties of Gaussians. In this work, we find that the data are close to Gaussian-distributed, with a single main peak and long tails. As such, we fit the widest Gaussian that is consistent with a Gaussian fit to the bulk of the distribution, and only fit a single Gaussian component. In this way we are separating Gaussian signal from non-Gaussian foregrounds, but not attempting to separate Gaussian 21cm signal and Gaussian foregrounds.

\section{Methods}\label{sec:methods}
We follow a similar approach as outlined in \citet{kde2019}. We present a brief overview here and refer the reader to the earlier work for more details. Individual observations from the MWA comprise 2-min of data with 0.5s/10~kHz or 2s/40~kHz native sampling (depending on the phase of the array). The data are calibrated and combined according to prescription in \citet{nunhokee2024}. Briefly, the data are calibrated using Hyperdrive to obtain per-channel Direction Independent complex gains for each tile, and have a visibility-based source model directly subtracted from them. The calibration model contains 8,000 point sources and extended source components, and the subtraction model contains the brightest 4,000. Data are also averaged to 8s/80~kHz for the remainder of the analysis. After calibration, each observation passes through a set of quality metrics that assess the health of the telescope, RFI conditions, and calibration goodness-of-fit \citep{nunhokee23}. Observations that pass are then gridded onto a common ${uv}$-plane using the CHIPS pipeline \citep{trott16} with a Blackman Harris gridding kernel. A grid is produced for each frequency channel, yielding a complex-valued cube of gridded visibilities with dimensions $(u,v,\nu)$. In order to form power spectra with no noise power bias, data are time-interleaved to produce two sets with different noise realisations. In practice, the two sets are "totals" (whereby each set of two consecutive timesteps is added), and "differences" (whereby each set of two consecutive timesteps is differenced). The cross power spectrum can extracted by subtraction of the differences power from the totals power. A real-valued weights cube is also retained for optimal weighting of data to the final power spectrum. Finally, each $uv$-cell is spectrally transformed along the frequency axis (using a weighted Fourier Transform) to yield a cube in wavenumber space $(u,v,\eta) \propto (k_x,k_y,k_z)$. At this point the analysis diverges from that of the normal power spectrum estimation.

%A KDE aims to construct an estimate for the distribution function of identically distributed data, using the data themselves. The kernel is the underlying functional form for determining the contribution of a given measurement to the overall estimate. For an estimate with minimal bias, the kernel is best to be matched to the true underlying distribution, although this is often unknown (hence the use of KDE).

Each cell of the $(u,v,\eta)$ cube consists of a realisation of the underlying data distribution on a given vector spatial scale. An estimate of the true underlying distribution can be obtained by constructing a histogram from the independent sample realisations (i.e., cells) that correspond to the scalar spatial scale, $k^2 = k_x^2+k_y^2+k_z^2$. The Kernel Density Estimator is a tool for this, and represents each sample in the histogram by a pre-defined kernel. Each data sample is thermal noise-dominated, rendering a Gaussian kernel to be a natural choice for our purposes. Moreover, the use of a KDE to construct the power spectrum necessitates a Gaussian kernel from which the variance can be equated to the power.

%For complex-valued visibility data from a radio interferometer with high spectral and temporal resolution (as is the case for EoR data), % This approach therefore naturally destroys any non-Gaussianity in the 21-cm signal, but may also allow for clean removal of non-Gaussian foreground contributions.

For a set of $N$ identically- and independently-distributed (iid) data, $x_j$, an estimate of their distribution over abscissa $x$ can be found via:
\begin{equation}
\hat{f}(x) = \frac{1}{Nh}\displaystyle\sum_{j=1}^N K\left( \frac{x_j-x}{h} \right),
\end{equation}
where the summation extends over the iid data, $K()$ denotes the compact kernel function, and $h$ is a scaling of the breadth of the kernel, for which an optimal value for Gaussian-distributed data is found to be \citep{silverman86}:
\begin{equation}
h \approx 1.06\sigma(N)^{-1/5},
\end{equation}
where $\sigma$ is the standard deviation of the data. For radio interferometric visibilities, for which the real and imaginary components contain an equal share of 21-cm signal and Gaussian noise measured in Janskys\footnote{Strictly, the signal here is in units of Jy.Hz, because the spectral transform from $\nu$ to $\eta$ space has already been performed. However, in practise, we retain the Jy unit and incorporate the spectral units in the final cosmological normalization (including the $\Delta\nu$ multiplication).}, $S$:
\begin{equation}
\hat{f}(S_i) = \frac{1}{N_{i}h}\displaystyle\sum_{j=1}^{N_i} K\left( \frac{S_{ij}(k_\bot,k_\parallel)-S}{h} \right),
\end{equation}
with
\begin{equation}
K\left( \frac{S_{i}(k_\bot,k_\parallel)-S}{h} \right) = \exp{-\frac{(S_{i}(k_\bot,k_\parallel)-S)^2}{2h^2}},
\end{equation}
and for $i \in [k_\bot,k_\parallel]$. We then connect the variance to the power and equate,
\begin{eqnarray}
&A_i\exp{\frac{-(S_{i}(k_\bot,k_\parallel)-S)^2}{2P_S(k_\bot,k_\parallel)}} \\\nonumber
&= \displaystyle\sum_{j=1}^{N_i} w_j\exp{\frac{-(S_{ij}(k_\bot,k_\parallel)-S)^2}{2h^2}},
\end{eqnarray}
where $P_S(k_\bot,k_\parallel)$ is the power measured in units of Jy$^2$, and $A$ is an amplitude that depends on the number of visibilities contributing to cell $i$, and is unused except to estimate the noise uncertainty. The factor $w_j$ denotes the weight of each gridded visibility and provides an optimal estimate of the histogram. In this work, we use an abscissa of flux density, $F$, with range [--150, 150]~Jy, and a resolution of 1~mJy, for each of the real and imaginary components of the gridded data. 150~Jy was empirically-determined by inspection of the data; other ranges of histogram were trialled and the results found to be robust to this choice (which captured all of the data that were not affected by RFI, and flagged). At $z=6.8$ this corresponds to a resolution of 57 mK~Mpc$^{1.5}$ (corresponding to a power resolution of 3,300 mK$^2$ Mpc$^3$), which over-samples the expected 21cm signal power in the EoR window by a factor of 50--100 (an expected signal level in these modes is 10$^5$ mK$^2$ Mpc$^3$ h$^{-3}$). The kernel width, $h$ is set to 3 times the abscissa resolution. Both the binning and $h$ values were tested to ensure robustness of the results.

The parameters of the Gaussians (amplitude, mean, variance) are fitted using the IDL \texttt{GAUSSFIT} routine. In this work, a separate estimate of $P_S$ is obtained for the real and imaginary components of the visibility, and for the totals and difference sets of gridded data. As such, the final cosmological power spectrum is obtained as,
\begin{equation}
    P = \frac{1}{4}\left( \sqrt{P_{\rm{tot},r}^2 + P_{\rm{tot},i}^2} - \sqrt{P_{\rm{diff},r}^2 + P_{\rm{diff},i}^2}  \right).
\end{equation}

As an alternative to the fitted Gaussian approach, we also use the histograms, $h(f)$, to estimate the variance via a basic Cumulative Distribution Function (CDF), such that:
\begin{equation}
    P_S = \frac{\displaystyle\sum_{f=0}^{N_f}h(f)F^2(f)}{\displaystyle\sum_{f=0}^{N_f}h(f)},
\end{equation}
where $N_f=300,000$ is the size of flux density abscissa, and $F(f)$ is the histogram flux density value at index $f$. In the case of a purely Gaussian distribution, the fitted Gaussian and CDF approaches should yield the same value for the variance. The CDF will be used as a "blind" benchmark statistic where any non-Gaussianity is ignored in the estimate, thereby producing results that are more akin to the normal CHIPS power spectrum estimator\footnote{Note that the results will not be identical, due to the CDF using the histogram of the values, and CHIPS using the coherently-gridded weights}. Throughout, the CDF, fitted histograms, and CHIPS power spectra use the exact same underlying datasets.

In addition to estimating the power separately from the NS and EW polarisations, we can combine the two histograms. The individual samples are incoherent realisations of the signal power, thereby allowing for data from different polarisations, and different observing fields (if available) to included in the same histogram. In this work, we consider the polarisations separately before combining them to a common estimate. In this work, we also study the behaviour across the real and imaginary components, and across redshift.

\section{Results\label{subsec:figures}}
\begin{figure}[ht!]
\plotone{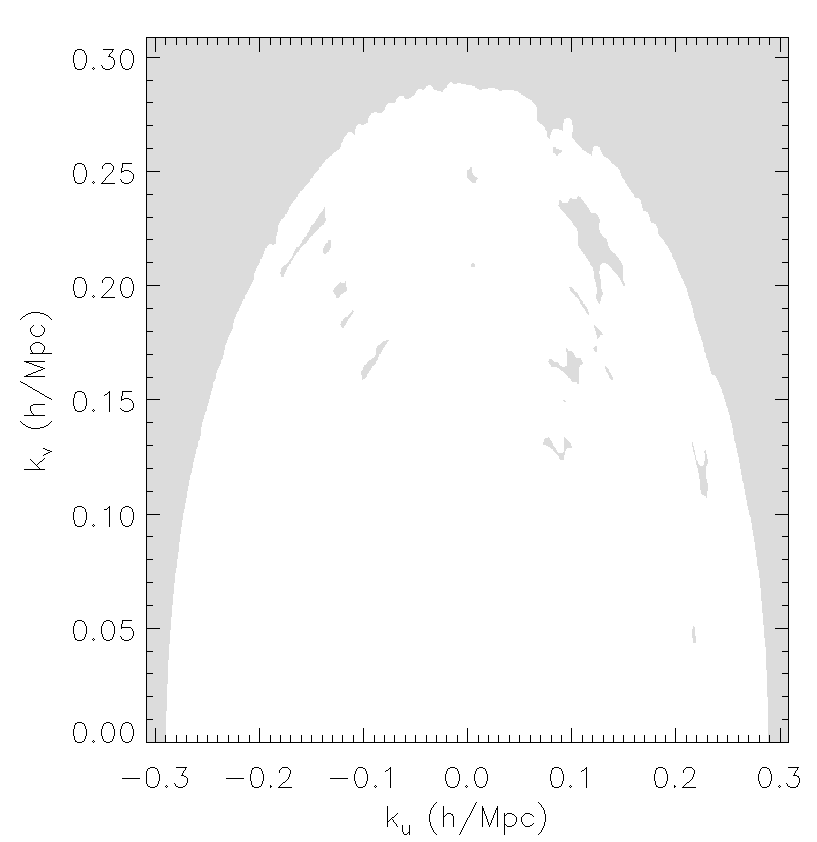}
\caption{Occupancy of measurements in the $uv$-plane for the 8,036 observation dataset, for a single frequency channel. The white-shaded region shows the parts of the plane where there are measurements available, demonstrating that complete $uv$-coverage is obtained for $k_\bot < 0.12 h$/Mpc. 
\label{fig:uv}}
\end{figure}

In this work all 8,036 observations (268 hours) are used that passed QA from \citet{nunhokee2024}. Figure \ref{fig:uv} shows the density of measurements in the $uv$-plane for the 8,036 observation dataset, for a single frequency channel. The white-shaded region shows the parts of the plane where there are measurements available, demonstrating that complete $uv$-coverage is obtained for $k_\bot < 0.12 h$/Mpc. This motivates the use of a KDE to be able to robustly reproduce the underlying data distribution.

\begin{figure*}[ht!]
\plottwo{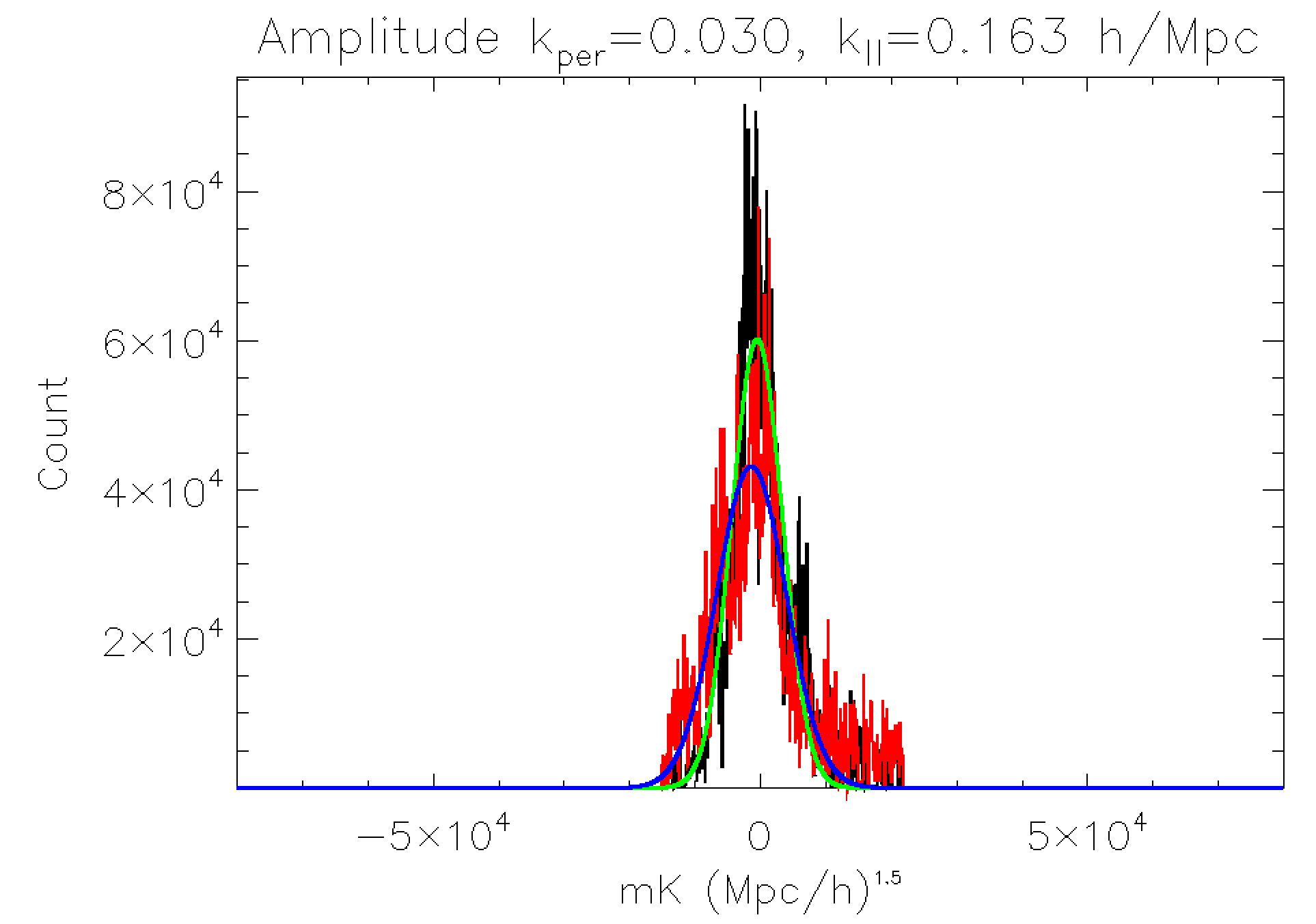}{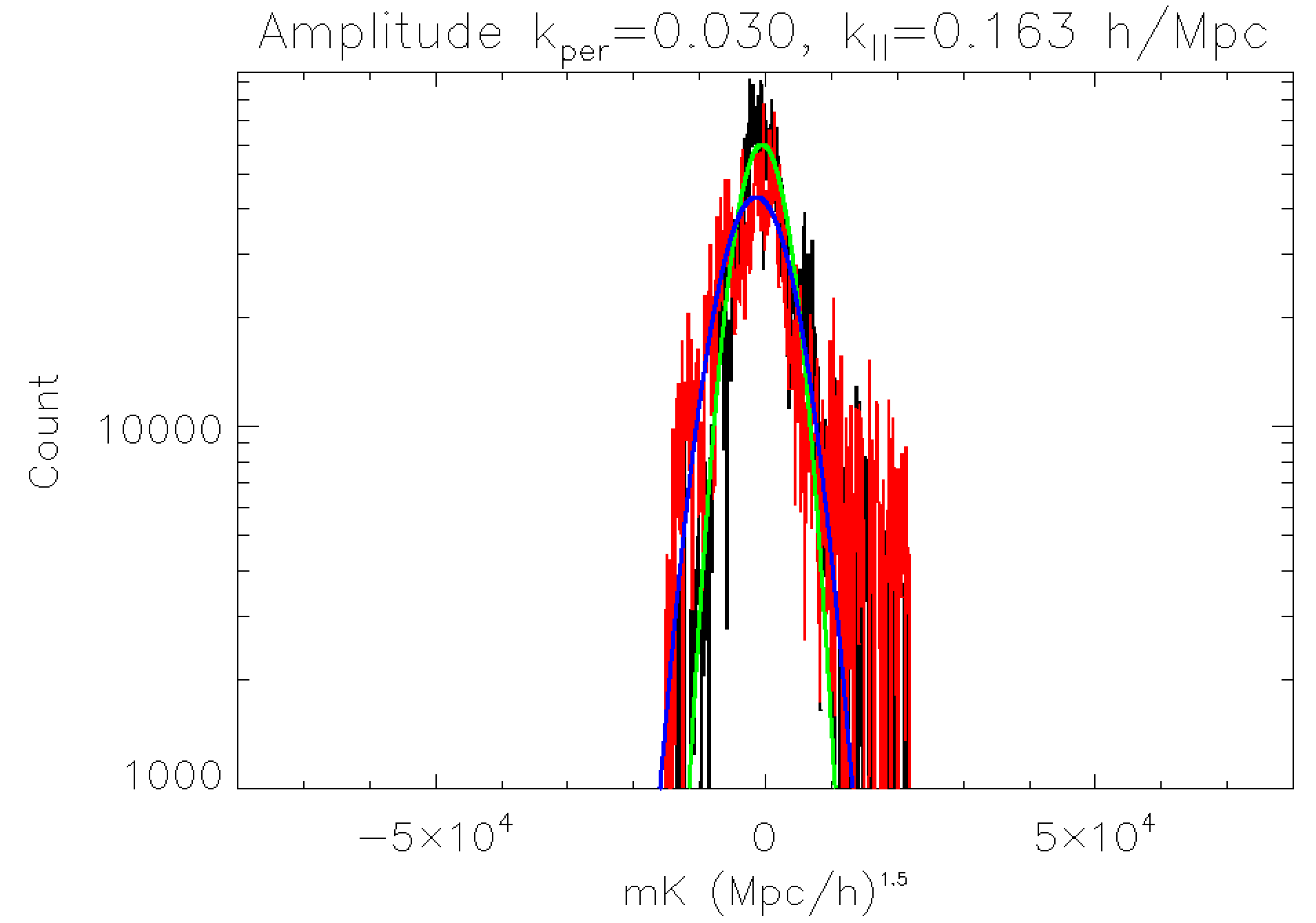}
\caption{(Left) Histograms of the real (red) and imaginary (black) components of the data in the cell at $k_\bot=0.03 h$/Mpc,  $k_\parallel=0.17 h$/Mpc and the EW polarisation, and their Gaussian best-fits (green, blue solid). (Right) Same, but using a logarithmic scale for the Count axis to highlight the outliers.
\label{fig:histos}}
\end{figure*}

\begin{figure*}[ht!]
\plotone{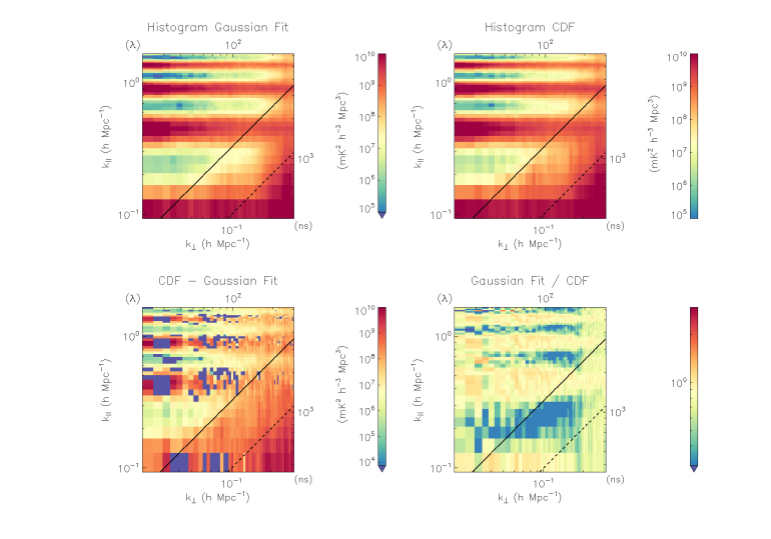}
\caption{2D power for the 8,036 high-band EoR0 observations in the EW polarisation. (Top-left) Power extracted from Gaussian fitting to distributions; (top-right) power extracted by integrating over full histogram CDFs; (bottom-left) Difference: Integrated power - Fitted power; (bottom-right) Ratio: Fitted power / Integrated power. 
\label{fig:2D_8036_EW}}
\end{figure*}

Figure \ref{fig:histos} shows examples of the histograms produced from the gridded data for a cell in the EoR window, at $k_\bot=0.03 h$/Mpc,  $k_\parallel=0.16 h$/Mpc, for data from $z=6.5$. This is a cell that shows some improvement between the CDF approach and the fitted Gaussian approach, and the reasons for this can be observed in the fits to the histograms and the tail of signal.
The corresponding 2D power spectrum across all cells is shown in Figure \ref{fig:2D_8036_EW}.
The ratio (bottom-left) demonstrates that foreground-dominated modes tend to produce similar results across the two estimation approaches, except for modes near the edge of the wedge (horizon power), and modes in the lower EoR window. In some cells, the ratio of power estimates is nearly three. These are modest improvements with respect to the overall power level, but will improve limits in the EoR window.

\begin{figure*}[ht!]
\plotone{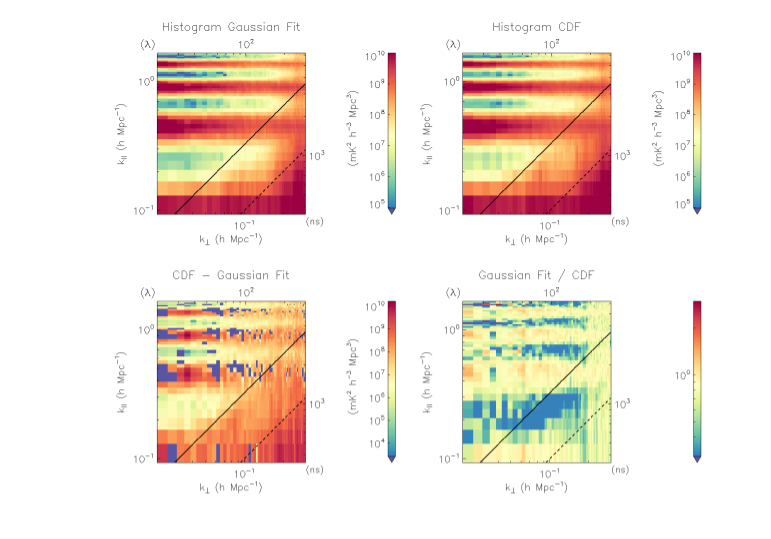}
\caption{2D power for the 8,036 high-band EoR0 observations in the NS polarisation. (Top-left) Power extracted from Gaussian fitting to distributions; (top-right) power extracted by integrating over full histogram CDFs; (bottom-left) Difference: Integrated power - Fitted power; (bottom-right) Ratio: Fitted power / Integrated power. 
\label{fig:2D_8036_NS}}
\end{figure*}

\begin{figure}[ht!]
\plotone{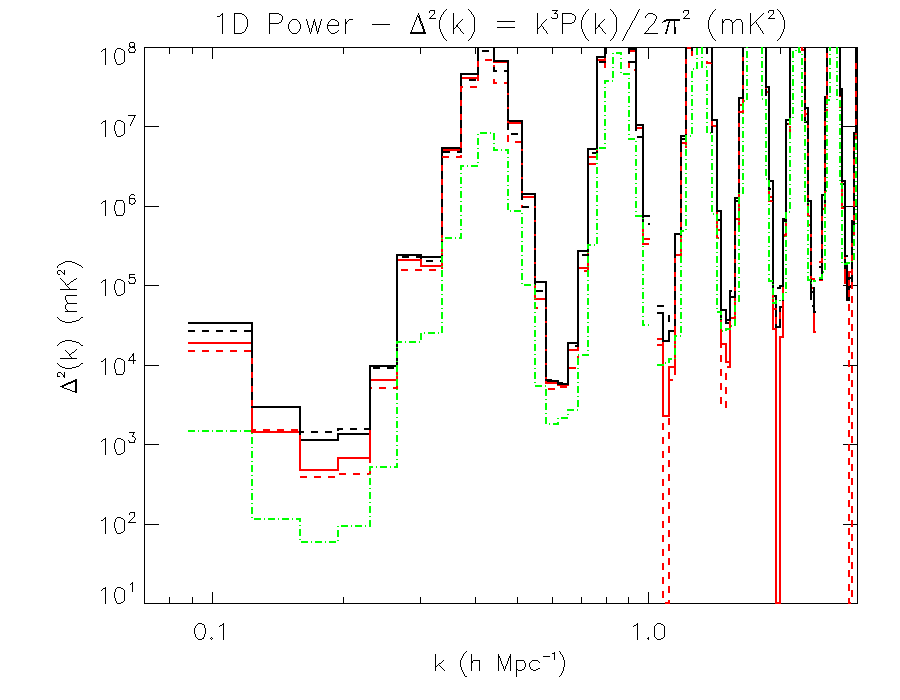}
\caption{1D power for the EW (solid) and NS (dashed) polarisations for 8,036 observations at $z=6.5$, and the noise level (green). The black lines denote the CDF approach, whereas the red denote the fitted Gaussians. 
\label{fig:1D_10_NS_EW}}
\end{figure}

\begin{figure*}[ht!]
\plotone{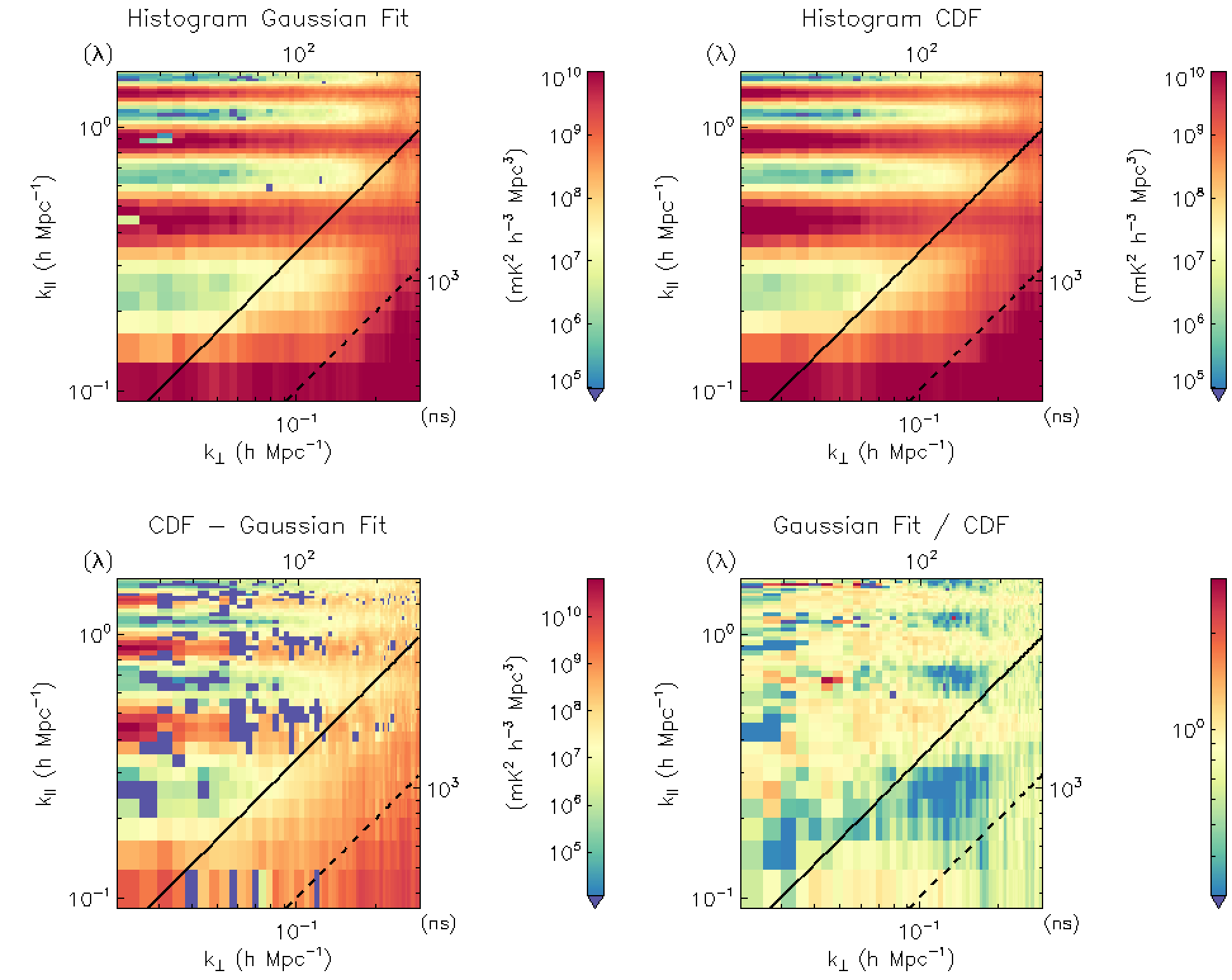}
\caption{2D power spectra from the fitted Gaussians and CDF approaches for 8,036 NS and EW visibilities at $z=6.5$. Similar behaviour is observed in the ratio as for the EW and NS separately.
\label{fig:NSEW_8036_2D}}
\end{figure*}

\begin{figure}[ht!]
\plotone{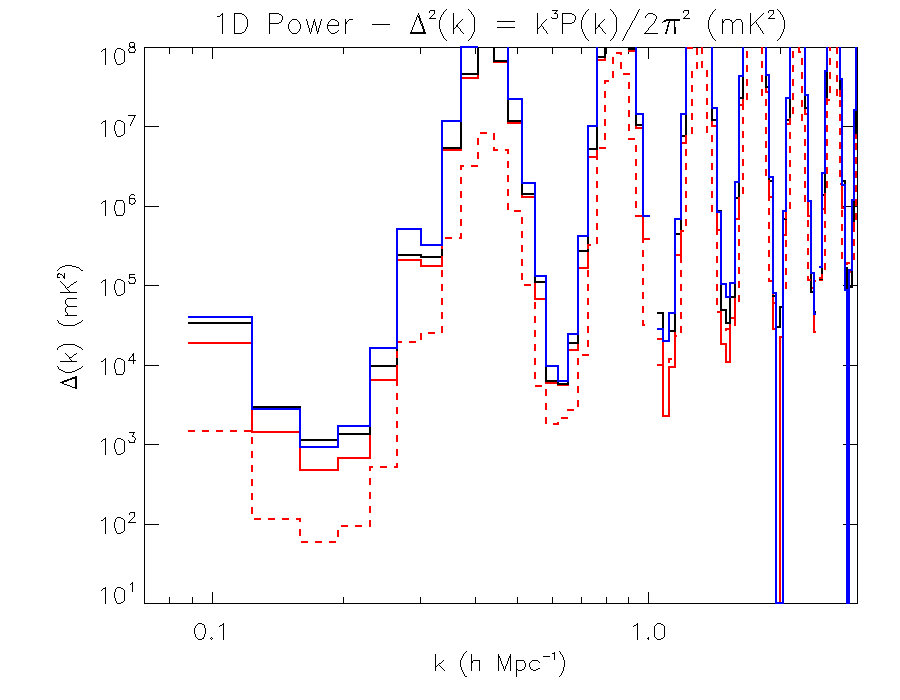}
\caption{Spherically-averaged power spectrum for the 8,036 observations at $z=6.5$ using both polarizations. Solid blue denotes the result from \citet{nunhokee2024} using a subset of 6,395 observations, solid black denotes the CDF-derived power, and solid red denotes the Gaussian fitted power. The dashed line is the 2$\sigma$ noise corresponding to the Gaussian-fitted power.
\label{fig:NSEW_8036_1D}}
\end{figure}

Similar analysis can be performed for the NS polarisation, in which there was more contamination seen in \citet{nunhokee2024}. The results are shown in Figure \ref{fig:2D_8036_NS} and demonstrate greater improvement than EW for $k_\bot < 0.02h$/Mpc.
The 1D power spectrum for both of those polarisations and $z=6.5$ is shown in Figure \ref{fig:1D_10_NS_EW}.
There is overall improvement in both polarisations for $k_\bot \sim 0.15-0.20h$/Mpc, but the improvement is greater for the NS polarisation. Note that both polarisations exhibit similar behaviour, as expected. The final step is to combine the two polarisations to produce NS+EW power spectra.
The resultant 2D and 1D spectra are shown in Figures \ref{fig:NSEW_8036_2D} and \ref{fig:NSEW_8036_1D}, respectively.
%% The "ht!" tells LaTeX to put the figure "here" first, at the "top" next
%% and to override the normal way of calculating a float position

%\begin{figure}[ht!]
%\plotone{Vis_histograms_kper6_kpa7.png}
%\caption{The cost for an author to publish an article has trended downward
%over time
%\label{fig:general}}
%\end{figure}

In table \ref{table:limits} we present the existing limits (where available) from \citet{nunhokee2024} and the improved limits of this work. 
\begin{table*}
\centering
\begin{tabular}{|c||c|c|c|c|c|c|c|}
\hline 
$z$ & $k$ (h/Mpc) & P$_{UL}$ (mK$^2$) & P$_{therm}$ (mK$^2$) & P$_{sample}$ (mK$^2$) & P$_{\rm CDF}$ (mK$^2$) & P$_{\rm orig}$ (mK$^2$)  & Pol. \\ 
\hline \hline
6.5 & 0.18 &  (23.0)$^2$ & (6.7)$^2$ & (6.1)$^2$ & (34.5)$^2$ & (30.2)$^2$  & EW \\
6.8 & 0.18 &  {\bf (28.4)$^2$} & (9.3)$^2$ & (6.9)$^2$ & (35.4)$^2$ & (31.3)$^2$  &  EW \\
7.0 & 0.21 &  (35.4)$^2$ & (14.2)$^2$ & (9.4)$^2$ & (40.8)$^2$ & (39.1)$^2$  & EW \\
\hline
6.5 & 0.18 &  {\bf (21.7)$^2$} & (8.4)$^2$ & (5.8)$^2$ & (39.0)$^2$ & (39.2)$^2$  & NS \\
6.8 & 0.18 &  (32.4)$^2$ & (9.3)$^2$ & (8.6)$^2$ & (86.5)$^2$ & (79.3)$^2$ & NS \\
7.0 & 0.21 &  (35.3)$^2$ & (14.2)$^2$ & (8.9)$^2$ & (61.0)$^2$  & (55.3)$^2$ & NS \\
\hline 
6.5 & 0.18 &  (23.0)$^2$ & (7.1)$^2$ & (6.4)$^2$ & (35.3)$^2$ & -- &  NS,EW \\
6.8 & 0.18 &  (30.5)$^2$ & (7.8)$^2$ & (8.4)$^2$ & (52.8)$^2$ & -- &  NS,EW \\
7.0 & 0.18 &  {\bf (34.9)$^2$} & (9.1)$^2$ & (10.3)$^2$ & (44.8)$^2$ & -- &  NS,EW \\
\hline
\end{tabular}
\caption{Best 2$\sigma$ upper limits for each redshift and each polarization, where both the thermal noise and sample variance are included in the error budget. "UL" denotes the upper limits from this work; "CDF" denotes the output of the Cumulative Distribution Function of the histograms from this work; "orig" denotes the limits from the same underlying dataset from \citet{nunhokee2024}. Bold-faced values are the best limits for each redshift.}\label{table:limits}
\end{table*} 
The best limit improvements are factors of 1.5 ($z=6.5$), 1.9 ($z=6.8$), and 2.0 ($z=7.0$). 

\section{Foreground sky simulation}
\begin{figure*}[ht!]
\plotone{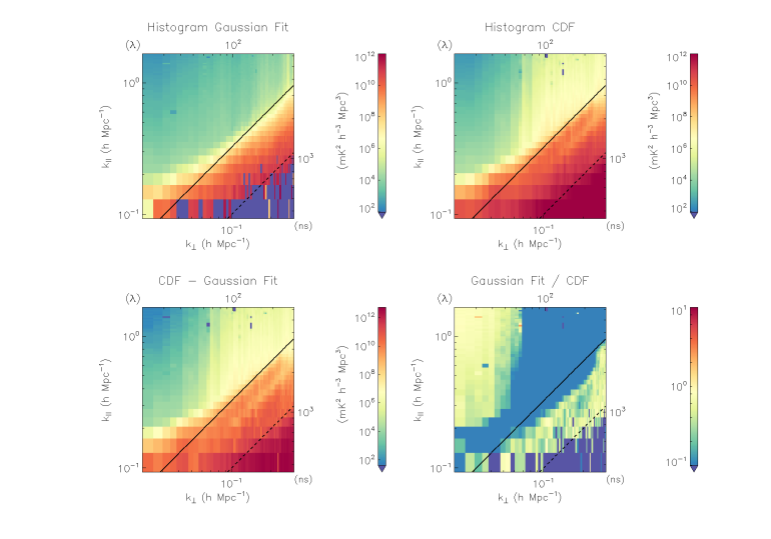}
\caption{2D power spectrum for the Gaussian fit (top-left), CDF (top-right), the fractional difference (bottom-left), and the ratio (bottom-right) for the EW polarisation data containing diffuse and compact emission. Note the different colour scale than other similar plots. The purple modes close to $k_\parallel =0$ are those not fitted by the algorithm and can be ignored.}
\label{fig:FG}
\end{figure*}

\begin{figure}[ht!]
\plotone{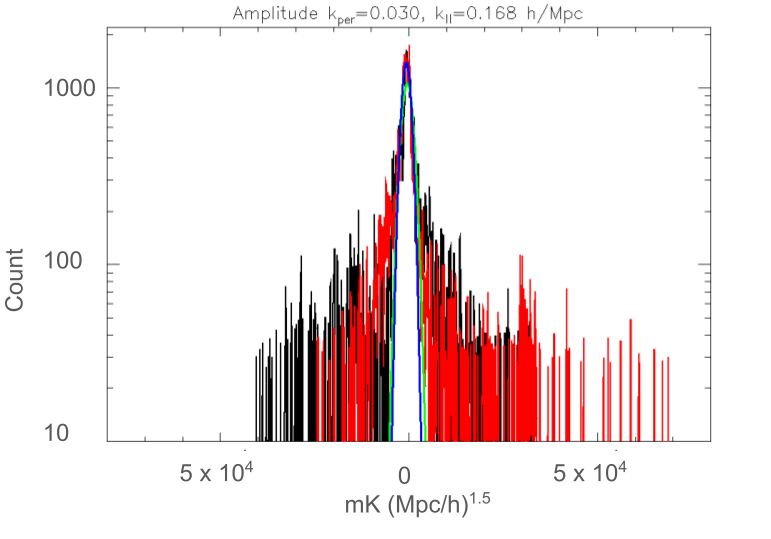}
\caption{The same $uv$ cell as used in Figure \ref{fig:histos} for the diffuse plus compact emission simulations. The clear non-Gaussian components can be observed, and look qualitatively similar to the data.}
\label{fig:FG_histo}
\end{figure}

To support the hypothesis that non-Gaussianity is due to foreground signal, observed through the telescope, we undertake realistic simulations of the diffuse (Galactic emission) and compact (point source, double, extended sources) sky, and observe them through the MWA. We use data simulated with the WODEN software\footnote{https://woden.readthedocs.io/en/latest/index.html} \citep{line22,line24}, and include a diffuse sky map derived from \citet{kriele22}, along with the compact sky model used for the MWA, comprising GLEAM \citep{gleam17}, GLEAM-X \citep{gleam24} and LoBES \citep{lobes21}, and with additional modelling from \citet{cook22} and \citet{line22}. The simulation spans five observations across the LST range of the observations, and are noise-free. These simulated data have uv-coverage that is complete for $k_\bot < 0.1h$/Mpc, and $>$90\% complete radially, and 100\% complete azimuthally for $k_\bot < 0.2h$/Mpc. Figure \ref{fig:FG} shows the same set of cylindrical power spectra as were presented for the data. Note that the colour bar scales have changed to represent the amplitudes of the power and ratios.
Aside from a few modes at $k_\parallel = 0$, which have not converged in their Gaussian fitting (due to the breadth of the distributions), the simulations demonstrate similar behaviour to the data near the edge of the wedge and in the EoR window; i.e., that there are non-Gaussian foregrounds that reside in this part of the parameter space, and that the Gaussian fitting therefore estimates a smaller variance than a blind CDF approach. Figure \ref{fig:FG_histo} shows the same $uv$ cell as used in Figure \ref{fig:histos} for these simulations. The clear non-Gaussian components can be observed, and look qualitatively similar to the data.
These simulations therefore support the hypothesis that leaking non-Gaussian foregrounds are contaminating the results.

\section{21cm Validation}
The techniques presented here intend to reproduce the Gaussian component of data. We can compare a simulation of the 21cm signal with both methods to validate the approach. We again use data simulated with the WODEN software. The underlying 21cm simulation was produced to match the MWA's large field-of-view and presented in \citet{greig22}. It uses a large 21cmFAST cube \citep{mesinger07,mesinger11} sampled at the MWA spectral resolution. Thirty observations, covering the pointings and LSTs to match those for the data used here, were simulated and combined into a single dataset.
\begin{figure*}[ht!]
\plotone{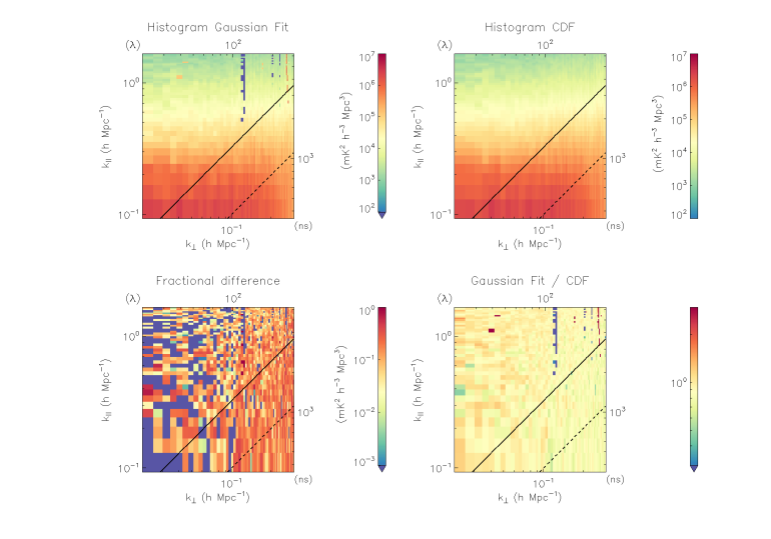}
\caption{2D power spectrum of 21cm signal for the Gaussian fit (top-left), CDF (top-right), the fractional difference (bottom-left), and the ratio (bottom-right) for the EW polarisation data. The purple modes are those not fitted by the algorithm and can be ignored.}
\label{fig:21cm_2d}
\end{figure*}

\begin{figure}[ht!]
\plotone{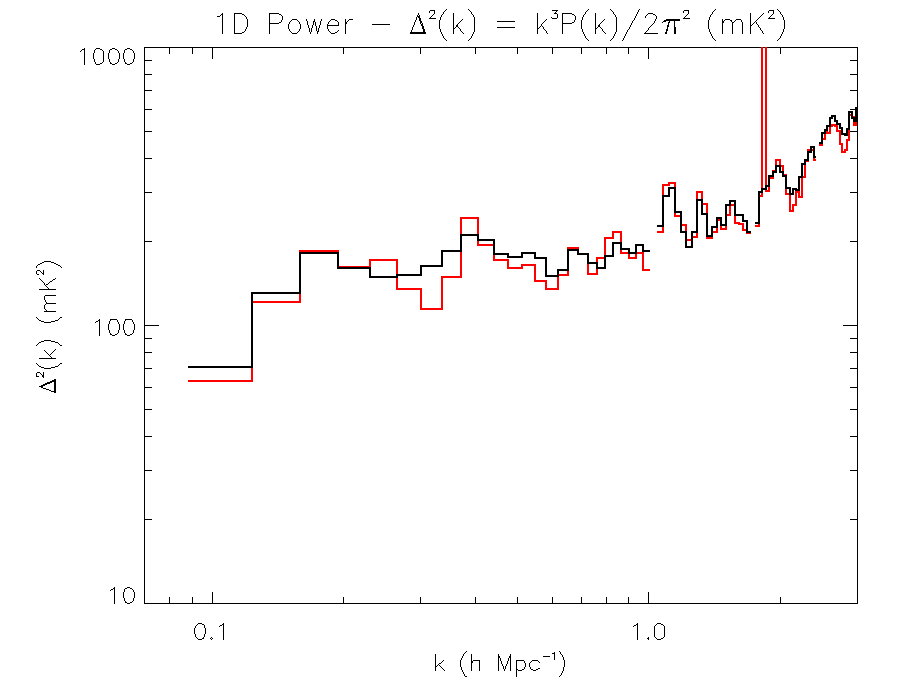}
\caption{Gaussian fit (red) and CDF (black) of a 21cm simulation using the WODEN software tool and encompassing the same range of pointings as the data. The two fits match well across wavenumber. 
\label{fig:21cm_1d}}
\end{figure}

Figure \ref{fig:21cm_2d} shows the 2D power spectrum for the Gaussian fit (top-left), CDF (top-right), the fractional difference (bottom-left), and the ratio (bottom-right) for the EW polarisation data. Figure \ref{fig:21cm_1d} shows the corresponding 1D power spectrum over $k_\bot = 0.02-0.035$~h/Mpc. Unlike the actual EoR0 data, these simulated 21cm visibilities show a high degree of Gaussianity, as expected, and no systematic difference between the histogram-fit and CDF approaches. The mean ratio in the bottom-right plot is 0.96 across the full parameter space, and exceeds unity in the EoR window (Gaussian fitting slightly over-estimating the power).

\begin{figure*}[ht!]
\plotone{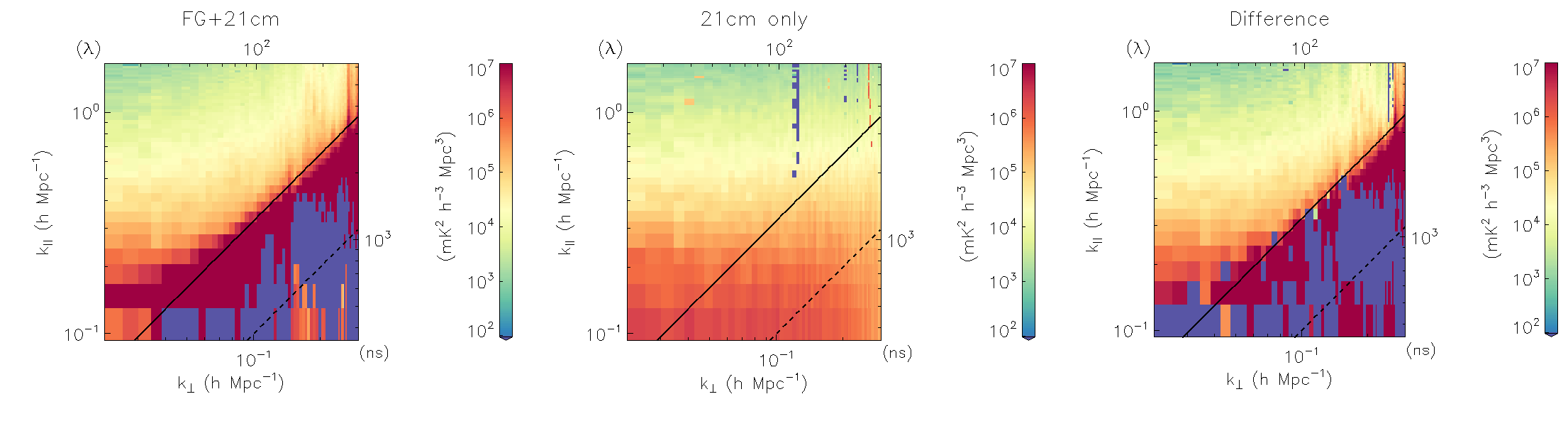}
\caption{Full 21cm $+$ compact and diffuse foregrounds simulation (left), the 21cm-only simulation (centre), and the difference (right), showing recovery of the 21cm signal in the EoR window, even in the presence of strong foregrounds. 
\label{fig:sim_both}}
\end{figure*}

The final check is to extend this analysis to include 21cm signal and diffuse and compact foregrounds in the simulation, and to show that the 21cm signal can be recovered in the EoR window. The results of that simulation are shown in Figure \ref{fig:sim_both}, where the full simulation (left) and the 21cm-only simulation (centre), are differenced (right) to show recovery of the 21cm signal in the EoR window. 
The consistency of the two sets of simulations demonstrates that Gaussian-like 21~cm signal can be recovered adequately. 

We chose the $z$=6.5 range of the Greig simulation for the validation, to match with the redshift at which we had the most competitive results. The neutral fraction in that box is 40-45\% \citep{cook24}. The work of \citet{cook24}, which post-processed these boxes to study the power spectrum and skew spectrum shows non-zero skew at these late redshifts. In configuration space, tilde half of the voxels will have zero contrast temperature. The other half will encode the fluctuations in the remaining signal. In Fourier space, we are performing a weighted sum over the voxels, where the weight is a sinusoidal function. The average of the box is not measured (contained in the auto-correlation, which we discard), and each Fourier mode that a histogram is produced for would be a sum over signal and zeros. In configuration space, one may guess that the distribution would be a spike at T=0, embedded within a Gaussian distribution. The FT of this is a flat line (equal power across $k$) summed with a Gaussian by the linearity of the Fourier Transform, leading to some small non-Gaussianity on top of a Gaussian distribution. The degree of non-Gaussianity we estimated based on the work of \citet{cook24} and \citet{ma23} suggested that this broadening would not be measurable with the precision of our data.

\section{Discussion and Conclusions}
The realistic simulations produced by WODEN are crucial to support use of this approach, and also to target the source of the non-Gaussianity. At the level of the signal we can probe with 268 hours of data, any non-Gaussianity in the 21~cm signal itself is insignificant. However, even if there was substantial non-Gaussianity, the power spectrum statistic is not intended to measure it, but instead extract the second moment of the distribution. While the 21~cm simulations show no measurable deviation from Gaussianity, the diffuse $+$ compact emission simulations demonstrate clear non-Gaussianity in foreground dominated modes, but also in the modes close to the edge of the wedge, and in the EoR window. These modes are the ones where we find the most improvement in the 268 hours of MWA data, and therefore are used to derive the improved upper limits.

The comparison of the results from the two polarizations is also interesting. Both use the same underlying observations, but have slightly different responses to the sky. The results from \citet{nunhokee2024} showed a large difference between the polarizations, with the NS polarization, which has a relatively larger response to the Galactic Plane emission near the Western horizon, yielding poorer performance. Galactic emission, being spatially-localized and non-stochastic, would be expected to produce some non-Gaussianity (compare with point sources distributed across the sky, which would produce a Gaussian response according to the central limit theorem). Under this hypothesis, fitting a Gaussian model to the distributions for NS may avoid a lot of that Galactic emission and produce results that were more similar to the EW polarization. Indeed, the results from Table \ref{table:limits} show much more consistency between polarizations for each redshift than the CDF or results from \citet{nunhokee2024}, particularly for $z=6.5, 7.0$.
\begin{figure*}[ht!]
\plotone{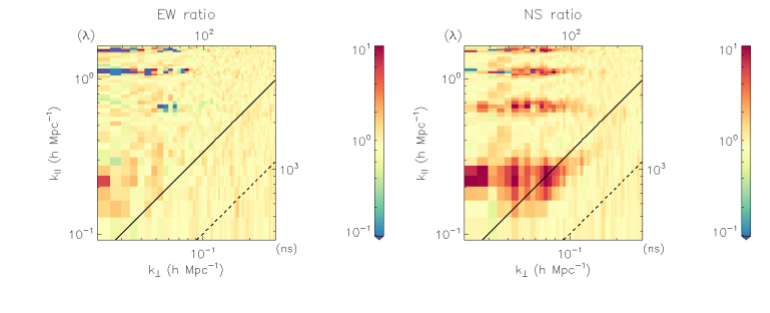}
\caption{The ratio of the CDF-derived power to the Gaussian fitted power for the EW (left) and NS (right) polarizations at $z=6.8$. There is clear improvement shown in the NS polarization relative to the EW.}
\label{fig:ratios_band12}
\end{figure*}
This can be observed in Figure \ref{fig:ratios_band12}, which shows the ratio of the CDF-derived power to the Gaussian fitted power for the EW (left) and NS (right) polarizations at $z=6.8$. There is clear improvement shown in the NS polarization relative to the EW.

We are proceeding from the intuition that any additional Gaussian or non-Gaussian component will always broaden the distribution, and so we are safe to make that assumption as a conservative estimate. This work is motivated by the intuition that any Gaussian or non-Gaussian foreground signal will work to broaden the distribution of the data. This will not be the case if somehow the foregrounds are anti-correlated with the 21cm signal and works to narrow the distribution. In addition to there being no physical motivation for this, the work undertaken in \citet{kde2019} looked at the performance of this histogram approach to both the EoR0 and EoR1 fields, finding similar behaviour for both. Anticorrelation of 21cm signal and foregrounds would need to conspire to produce this effect in both of these fields, adding weight to this not being the explanation. Additionally, the results of the 21cm validation are also suggestive that this conspiracy is not at play.

The approach used in this research will work well with other foreground-dominated datasets, and will reach its limit in the regime where a detection of the 21~cm second moment-only is possible (higher-order statistics are not yet detectable). In those future studies with increased sensitivity currently unavailable to us, an exploration of the distribution of data will provide more information about the early Universe than the power spectrum alone, and provides the large-scale cosmic statistical information to complement the more localised tomographic studies.

%% 
%% Also note that the akcnowlodgment environment does not support long amounts of text. If you have a lot of people and institutions to acknowledge, do not use this command. Instead, create a new \section{Acknowledgments}.
\begin{acknowledgments}
We thank Adrian Liu, as the reviewer of this manuscript, for his rigorous examination of the work, and its consequent improvement. This research was partly supported by the Australian Research Council Centre of Excellence for All Sky Astrophysics in 3 Dimensions (ASTRO 3D), through project number CE170100013. The International Centre for Radio Astronomy Research (ICRAR) is a Joint Venture of Curtin University and The University of Western Australia, funded by the Western Australian State government. This scientific work uses data obtained from \textit{Inyarrimanha Ilgari Bundara} / the Murchison Radio-astronomy Observatory. We acknowledge the Wajarri Yamaji People as the Traditional Owners and native title holders of the Observatory site. Establishment of CSIRO’s Murchison Radio-astronomy Observatory is an initiative of the Australian Government, with support from the Government of Western Australia and the Science and Industry Endowment Fund. Support for the operation of the MWA is provided by the Australian Government (NCRIS), under a contract to Curtin University administered by Astronomy Australia Limited. This work was supported by resources provided by the Pawsey Supercomputing Research Centre with funding from the Australian Government and the Government of Western Australia. JCP acknowledges support from the U.S. National Science Foundation grant \#2106510. MFM, BH, and EL acknowledge support from the U.S. National Science Foundation grant \#2107538.
\end{acknowledgments}


\vspace{5mm}
\facilities{Murchison Widefield Array}


\software{IDL, WODEN \citep{line24}
          }



\bibliography{sample631,references}

\begin{thebibliography}{}
\expandafter\ifx\csname natexlab\endcsname\relax\def\natexlab#1{#1}\fi
\providecommand{\url}[1]{\href{#1}{#1}}
\providecommand{\dodoi}[1]{doi:~\href{http://doi.org/#1}{\nolinkurl{#1}}}
\providecommand{\doeprint}[1]{\href{http://ascl.net/#1}{\nolinkurl{http://ascl.net/#1}}}
\providecommand{\doarXiv}[1]{\href{https://arxiv.org/abs/#1}{\nolinkurl{https://arxiv.org/abs/#1}}}

\bibitem[{{Acharya} {et~al.}(2024){Acharya}, {Mertens}, {Ciardi}, {Ghara}, {Koopmans}, \& {Zaroubi}}]{acharya24}
{Acharya}, A., {Mertens}, F., {Ciardi}, B., {et~al.} 2024, \mnras, 534, L30, \dodoi{10.1093/mnrasl/slae078}

\bibitem[{{Berkhout} {et~al.}(2024){Berkhout}, {Jacobs}, {Abdurashidova}, {Adams}, {Aguirre}, {Alexander}, {Ali}, {Baartman}, {Balfour}, {Beardsley}, {Bernardi}, {Billings}, {Bowman}, {Bradley}, {Bull}, {Burba}, {Byrne}, {Carey}, {Carilli}, {Chen}, {Cheng}, {Choudhuri}, {DeBoer}, {de Lera Acedo}, {Dexter}, {Dillon}, {Dynes}, {Eksteen}, {Ely}, {Ewall-Wice}, {Fagnoni}, {Fritz}, {Furlanetto}, {Gale-Sides}, {Garsden}, {Gehlot}, {Ghosh}, {Glendenning}, {Gorce}, {Gorthi}, {Greig}, {Grobbelaar}, {Halday}, {Hazelton}, {Hewitt}, {Hickish}, {Huang}, {Josaitis}, {Julius}, {Kariseb}, {Kern}, {Kerrigan}, {Kim}, {Kittiwisit}, {Kohn}, {Kolopanis}, {Lanman}, {La Plante}, {Liu}, {Loots}, {Ma}, {Edward MacMahon}, {Malan}, {Malgas}, {Malgas}, {Marero}, {Martinot}, {Mesinger}, {Molewa}, {Morales}, {Mosiane}, {Murray}, {Neben}, {Nikolic}, {Nunhokee}, {Nuwegeld}, {Parsons}, {Pascua}, {Patra}, {Pieterse}, {Qin}, {Rath}, {Razavi-Ghods}, {Riley}, {Robnett}, {Rosie}, {Santos}, {Sims}, {Singh}, {Storer}, {Swarts}, {Tan}, {Thyagarajan},
  {van Wyngaarden}, {Williams}, {Zheng}, \& {Xu}}]{berkhout24}
{Berkhout}, L.~M., {Jacobs}, D.~C., {Abdurashidova}, Z., {et~al.} 2024, \pasp, 136, 045002, \dodoi{10.1088/1538-3873/ad3122}

\bibitem[{{Bowman} {et~al.}(2013){Bowman}, {Cairns}, {Kaplan}, {Murphy}, {Oberoi}, \& {others}}]{bowman13_mwascience}
{Bowman}, J.~D., {Cairns}, I., {Kaplan}, D.~L., {et~al.} 2013, PASA, 30, 31, \dodoi{10.1017/pas.2013.009}

\bibitem[{{Boyett} {et~al.}(2024){Boyett}, {Trenti}, {Leethochawalit}, {Calabr{\'o}}, {Metha}, {Roberts-Borsani}, {Dalmasso}, {Yang}, {Santini}, {Treu}, {Jones}, {Henry}, {Mason}, {Morishita}, {Nanayakkara}, {Roy}, {Wang}, {Fontana}, {Merlin}, {Castellano}, {Paris}, {Brada{\v{c}}}, {Malkan}, {Marchesini}, {Mascia}, {Glazebrook}, {Pentericci}, {Vanzella}, \& {Vulcani}}]{boyett24}
{Boyett}, K., {Trenti}, M., {Leethochawalit}, N., {et~al.} 2024, Nature Astronomy, 8, 657, \dodoi{10.1038/s41550-024-02218-7}

\bibitem[{{Cook} {et~al.}(2024){Cook}, {Balu}, {Greig}, {Trott}, {Line}, {Qin}, \& {Wyithe}}]{cook24}
{Cook}, J.~H., {Balu}, S., {Greig}, B., {et~al.} 2024, \mnras, 529, 2734, \dodoi{10.1093/mnras/stae593}

\bibitem[{{Cook} {et~al.}(2022){Cook}, {Trott}, \& {Line}}]{cook22}
{Cook}, J.~H., {Trott}, C.~M., \& {Line}, J.~L.~B. 2022, \mnras, 514, 790, \dodoi{10.1093/mnras/stac1330}

\bibitem[{{DeBoer} {et~al.}(2017){DeBoer}, {Parsons}, {Aguirre}, {Alexander}, {Ali}, {Beardsley}, {Bernardi}, {Bowman}, {Bradley}, {Carilli}, {Cheng}, {de Lera Acedo}, {Dillon}, {Ewall-Wice}, {Fadana}, {Fagnoni}, {Fritz}, {Furlanetto}, {Glendenning}, {Greig}, {Grobbelaar}, {Hazelton}, {Hewitt}, {Hickish}, {Jacobs}, {Julius}, {Kariseb}, {Kohn}, {Lekalake}, {Liu}, {Loots}, {MacMahon}, {Malan}, {Malgas}, {Maree}, {Martinot}, {Mathison}, {Matsetela}, {Mesinger}, {Morales}, {Neben}, {Patra}, {Pieterse}, {Pober}, {Razavi-Ghods}, {Ringuette}, {Robnett}, {Rosie}, {Sell}, {Smith}, {Syce}, {Tegmark}, {Thyagarajan}, {Williams}, \& {Zheng}}]{deboer16}
{DeBoer}, D.~R., {Parsons}, A.~R., {Aguirre}, J.~E., {et~al.} 2017, \pasp, 129, 045001, \dodoi{10.1088/1538-3873/129/974/045001}

\bibitem[{{Ellingson} {et~al.}(2009){Ellingson}, {Clarke}, {Cohen}, {Craig}, {Kassim}, {Pihlstrom}, {Rickard}, \& {Taylor}}]{ellingson09}
{Ellingson}, S.~W., {Clarke}, T.~E., {Cohen}, A., {et~al.} 2009, IEEE Proceedings, 97, 1421, \dodoi{10.1109/JPROC.2009.2015683}

\bibitem[{{Fronenberg} \& {Liu}(2024)}]{fronenburg24}
{Fronenberg}, H., \& {Liu}, A. 2024, \apj, 975, 222, \dodoi{10.3847/1538-4357/ad77cc}

\bibitem[{{Fujimoto} {et~al.}(2023){Fujimoto}, {Wang}, {Weaver}, {Kokorev}, {Atek}, {Bezanson}, {Labbe}, {Brammer}, {Greene}, {Chemerynska}, {Dayal}, {de Graaff}, {Furtak}, {Oesch}, {Setton}, {Price}, {Miller}, {Williams}, {Whitaker}, {Zitrin}, {Cutler}, {Leja}, {Pan}, {Coe}, {van Dokkum}, {Feldmann}, {Fudamoto}, {Goulding}, {Khullar}, {Marchesini}, {Maseda}, {Nanayakkara}, {Nelson}, {Smit}, {Stefanon}, \& {Weibel}}]{fujimoto23}
{Fujimoto}, S., {Wang}, B., {Weaver}, J., {et~al.} 2023, arXiv e-prints, arXiv:2308.11609, \dodoi{10.48550/arXiv.2308.11609}

\bibitem[{{Greig} {et~al.}(2022){Greig}, {Wyithe}, {Murray}, {Mutch}, \& {Trott}}]{greig22}
{Greig}, B., {Wyithe}, J. S.~B., {Murray}, S.~G., {Mutch}, S.~J., \& {Trott}, C.~M. 2022, \mnras, 516, 5588, \dodoi{10.1093/mnras/stac2506}

\bibitem[{{Harikane} {et~al.}(2023){Harikane}, {Ouchi}, {Oguri}, {Ono}, {Nakajima}, {Isobe}, {Umeda}, {Mawatari}, \& {Zhang}}]{harikane23}
{Harikane}, Y., {Ouchi}, M., {Oguri}, M., {et~al.} 2023, \apjs, 265, 5, \dodoi{10.3847/1538-4365/acaaa9}

\bibitem[{{HERA Collaboration} {et~al.}(2023){HERA Collaboration}, {Abdurashidova}, {Adams}, {Aguirre}, {Alexander}, {Ali}, {Baartman}, {Balfour}, {Barkana}, {Beardsley}, {Bernardi}, {Billings}, {Bowman}, {Bradley}, {Breitman}, {Bull}, {Burba}, {Carey}, {Carilli}, {Cheng}, {Choudhuri}, {DeBoer}, {de Lera Acedo}, {Dexter}, {Dillon}, {Ely}, {Ewall-Wice}, {Fagnoni}, {Fialkov}, {Fritz}, {Furlanetto}, {Gale-Sides}, {Garsden}, {Glendenning}, {Gorce}, {Gorthi}, {Greig}, {Grobbelaar}, {Halday}, {Hazelton}, {Heimersheim}, {Hewitt}, {Hickish}, {Jacobs}, {Julius}, {Kern}, {Kerrigan}, {Kittiwisit}, {Kohn}, {Kolopanis}, {Lanman}, {La Plante}, {Lewis}, {Liu}, {Loots}, {Ma}, {MacMahon}, {Malan}, {Malgas}, {Malgas}, {Maree}, {Marero}, {Martinot}, {McBride}, {Mesinger}, {Mirocha}, {Molewa}, {Morales}, {Mosiane}, {Mu{\~n}oz}, {Murray}, {Nagpal}, {Neben}, {Nikolic}, {Nunhokee}, {Nuwegeld}, {Parsons}, {Pascua}, {Patra}, {Pieterse}, {Qin}, {Razavi-Ghods}, {Robnett}, {Rosie}, {Santos}, {Sims}, {Singh}, {Smith}, {Swarts}, {Tan},
  {Thyagarajan}, {Wilensky}, {Williams}, {van Wyngaarden}, \& {Zheng}}]{hera23}
{HERA Collaboration}, {Abdurashidova}, Z., {Adams}, T., {et~al.} 2023, \apj, 945, 124, \dodoi{10.3847/1538-4357/acaf50}

\bibitem[{{Hurley-Walker} {et~al.}(2017){Hurley-Walker}, {Callingham}, {Hancock}, {Franzen}, {Hindson}, {Kapi{\'n}ska}, {Morgan}, {Offringa}, {Wayth}, {Wu}, {Zheng}, {Murphy}, {Bell}, {Dwarakanath}, {For}, {Gaensler}, {Johnston-Hollitt}, {Lenc}, {Procopio}, {Staveley-Smith}, {Ekers}, {Bowman}, {Briggs}, {Cappallo}, {Deshpande}, {Greenhill}, {Hazelton}, {Kaplan}, {Lonsdale}, {McWhirter}, {Mitchell}, {Morales}, {Morgan}, {Oberoi}, {Ord}, {Prabu}, {Shankar}, {Srivani}, {Subrahmanyan}, {Tingay}, {Webster}, {Williams}, \& {Williams}}]{gleam17}
{Hurley-Walker}, N., {Callingham}, J.~R., {Hancock}, P.~J., {et~al.} 2017, \mnras, 464, 1146, \dodoi{10.1093/mnras/stw2337}

\bibitem[{{Jacobs} {et~al.}(2016){Jacobs}, {Hazelton}, {Trott}, {Dillon}, {Pindor}, {Sullivan}, {Pober}, {Barry}, {Beardsley}, {Bernardi}, {Bowman}, {Briggs}, {Cappallo}, {Carroll}, {Corey}, {de Oliveira-Costa}, {Emrich}, {Ewall-Wice}, {Feng}, {Gaensler}, {Goeke}, {Greenhill}, {Hewitt}, {Hurley-Walker}, {Johnston-Hollitt}, {Kaplan}, {Kasper}, {Kim}, {Kratzenberg}, {Lenc}, {Line}, {Loeb}, {Lonsdale}, {Lynch}, {McKinley}, {McWhirter}, {Mitchell}, {Morales}, {Morgan}, {Neben}, {Thyagarajan}, {Oberoi}, {Offringa}, {Ord}, {Paul}, {Prabu}, {Procopio}, {Riding}, {Rogers}, {Roshi}, {Udaya Shankar}, {Sethi}, {Srivani}, {Subrahmanyan}, {Tegmark}, {Tingay}, {Waterson}, {Wayth}, {Webster}, {Whitney}, {Williams}, {Williams}, {Wu}, \& {Wyithe}}]{jacobs16}
{Jacobs}, D.~C., {Hazelton}, B.~J., {Trott}, C.~M., {et~al.} 2016, \apj, 825, 114, \dodoi{10.3847/0004-637X/825/2/114}

\bibitem[{{Koopmans} {et~al.}(2015){Koopmans}, {Pritchard}, {Mellema}, {Aguirre}, {Ahn}, {Barkana}, {van Bemmel}, {Bernardi}, {Bonaldi}, {Briggs}, {de Bruyn}, {Chang}, {Chapman}, {Chen}, {Ciardi}, {Dayal}, {Ferrara}, {Fialkov}, {Fiore}, {Ichiki}, {Illiev}, {Inoue}, {Jelic}, {Jones}, {Lazio}, {Maio}, {Majumdar}, {Mack}, {Mesinger}, {Morales}, {Parsons}, {Pen}, {Santos}, {Schneider}, {Semelin}, {de Souza}, {Subrahmanyan}, {Takeuchi}, {Vedantham}, {Wagg}, {Webster}, {Wyithe}, {Datta}, \& {Trott}}]{koopmans15}
{Koopmans}, L., {Pritchard}, J., {Mellema}, G., {et~al.} 2015, Advancing Astrophysics with the Square Kilometre Array (AASKA14), 1.
\newblock \doarXiv{1505.07568}

\bibitem[{{Kriele} {et~al.}(2022){Kriele}, {Wayth}, {Bentum}, {Juswardy}, \& {Trott}}]{kriele22}
{Kriele}, M.~A., {Wayth}, R.~B., {Bentum}, M.~J., {Juswardy}, B., \& {Trott}, C.~M. 2022, \pasa, 39, e017, \dodoi{10.1017/pasa.2022.2}

\bibitem[{{Line}(2022)}]{line22}
{Line}, J. L.~B. 2022, Journal of Open Source Software, 7(69), 3976, \dodoi{10.21105/joss.03676}

\bibitem[{{Line} {et~al.}(2024){Line}, {Trott}, {Cook}, {Greig}, {Barry}, \& {Jordan}}]{line24}
{Line}, J.~L.~B., {Trott}, C.~M., {Cook}, J.~H., {et~al.} 2024, \pasa, 41, e067, \dodoi{10.1017/pasa.2024.31}

\bibitem[{{Liu} \& {Shaw}(2020)}]{liu20}
{Liu}, A., \& {Shaw}, J.~R. 2020, \pasp, 132, 062001, \dodoi{10.1088/1538-3873/ab5bfd}

\bibitem[{{Lynch} {et~al.}(2021){Lynch}, {Galvin}, {Line}, {Jordan}, {Trott}, {Chege}, {McKinley}, {Johnston-Hollitt}, \& {Tingay}}]{lobes21}
{Lynch}, C.~R., {Galvin}, T.~J., {Line}, J.~L.~B., {et~al.} 2021, \pasa, 38, e057, \dodoi{10.1017/pasa.2021.50}

\bibitem[{{Ma} \& {Peng}(2023)}]{ma23}
{Ma}, Q.-B., \& {Peng}, L. 2023, \mnras, 523, 640, \dodoi{10.1093/mnras/stad1447}

\bibitem[{{Mesinger} \& {Furlanetto}(2007)}]{mesinger07}
{Mesinger}, A., \& {Furlanetto}, S. 2007, \apj, 669, 663, \dodoi{10.1086/521806}

\bibitem[{{Mesinger} {et~al.}(2011){Mesinger}, {Furlanetto}, \& {Cen}}]{mesinger11}
{Mesinger}, A., {Furlanetto}, S., \& {Cen}, R. 2011, \mnras, 411, 955, \dodoi{10.1111/j.1365-2966.2010.17731.x}

\bibitem[{{Mu{\~n}oz} {et~al.}(2024){Mu{\~n}oz}, {Mirocha}, {Chisholm}, {Furlanetto}, \& {Mason}}]{munoz24}
{Mu{\~n}oz}, J.~B., {Mirocha}, J., {Chisholm}, J., {Furlanetto}, S.~R., \& {Mason}, C. 2024, \mnras, 535, L37, \dodoi{10.1093/mnrasl/slae086}

\bibitem[{{Nunhokee} {et~al.}(2024){Nunhokee}, {Null}, {Trott}, {Jordan}, {Line}, {Wayth}, \& {Barry}}]{nunhokee23}
{Nunhokee}, C.~D., {Null}, D., {Trott}, C.~M., {et~al.} 2024, arXiv e-prints, arXiv:2409.03232, \dodoi{10.48550/arXiv.2409.03232}

\bibitem[{{Nunhokee} {et~al.}(2025){Nunhokee}, {Null}, {Trott}, {Barry}, {Qin}, {Wayth}, {Line}, {Jordan}, {Pindor}, {Cook}, {Bowman}, {Chokshi}, {Ducharme}, {Elder}, {Guo}, {Hazelton}, {Hidayat}, {Ito}, {Jacobs}, {Jong}, {Kolopanis}, {Kunicki}, {Lilleskov}, {Morales}, {Pober}, {Selvaraj}, {Shi}, {Takahashi}, {Tingay}, {Webster}, {Yoshiura}, \& {Zheng}}]{nunhokee2024}
---. 2025, arXiv e-prints, arXiv:2505.09097, \dodoi{10.48550/arXiv.2505.09097}

\bibitem[{{Parsons} {et~al.}(2010){Parsons}, {Backer}, {Foster}, {Wright}, {Bradley}, {Gugliucci}, {Parashare}, {Benoit}, {Aguirre}, {Jacobs}, {Carilli}, {Herne}, {Lynch}, {Manley}, \& {Werthimer}}]{parsons10}
{Parsons}, A.~R., {Backer}, D.~C., {Foster}, G.~S., {et~al.} 2010, AJ, 139, 1468.
\newblock \doarXiv{0904.2334}

\bibitem[{{Patil} {et~al.}(2016){Patil}, {Yatawatta}, {Zaroubi}, {Koopmans}, {de Bruyn}, {Jeli{\'c}}, {Ciardi}, {Iliev}, {Mevius}, {Pandey}, \& {Gehlot}}]{patil16}
{Patil}, A.~H., {Yatawatta}, S., {Zaroubi}, S., {et~al.} 2016, \mnras, 463, 4317, \dodoi{10.1093/mnras/stw2277}

\bibitem[{{Ross} {et~al.}(2024){Ross}, {Hurley-Walker}, {Galvin}, {Venville}, {Duchesne}, {Morgan}, {An}, {G{\"u}rkan}, {Hancock}, {Heald}, {Johnston-Hollitt}, \& {White}}]{gleam24}
{Ross}, K., {Hurley-Walker}, N., {Galvin}, T.~J., {et~al.} 2024, \pasa, 41, e054, \dodoi{10.1017/pasa.2024.57}

\bibitem[{{Silverman}(1986)}]{silverman86}
{Silverman}, B.~W. 1986, {Density Estimation for Statistics and Data Analysis}

\bibitem[{{Tang} {et~al.}(2021){Tang}, {Stark}, {Chevallard}, {Charlot}, {Endsley}, \& {Congiu}}]{tang21}
{Tang}, M., {Stark}, D.~P., {Chevallard}, J., {et~al.} 2021, \mnras, 503, 4105, \dodoi{10.1093/mnras/stab705}

\bibitem[{{Tang} {et~al.}(2024){Tang}, {Stark}, {Ellis}, {Sun}, {Topping}, {Robertson}, {Tacchella}, {Arribas}, {Baker}, {Bhatawdekar}, {Boyett}, {Bunker}, {Charlot}, {Chen}, {Chevallard}, {Jones}, {Kumari}, {Lyu}, {Maiolino}, {Maseda}, {Saxena}, {Whitler}, {Williams}, {Willott}, \& {Witstok}}]{tang24}
{Tang}, M., {Stark}, D.~P., {Ellis}, R.~S., {et~al.} 2024, \mnras, 531, 2701, \dodoi{10.1093/mnras/stae1338}

\bibitem[{{Tingay} {et~al.}(2013){Tingay}, {Goeke}, {Bowman}, {Emrich}, \& {others}}]{tingay13_mwasystem}
{Tingay}, S.~J., {Goeke}, R., {Bowman}, J.~D., {Emrich}, D., \& {others}. 2013, PASA, 30, 7, \dodoi{10.1017/pasa.2012.007}

\bibitem[{{Trott} {et~al.}(2016){Trott}, {Pindor}, {Procopio}, {Wayth}, {Mitchell}, {McKinley}, {Tingay}, {Barry}, {Beardsley}, {Bernardi}, {Bowman}, {Briggs}, {Cappallo}, {Carroll}, {de Oliveira-Costa}, {Dillon}, {Ewall-Wice}, {Feng}, {Greenhill}, {Hazelton}, {Hewitt}, {Hurley-Walker}, {Johnston-Hollitt}, {Jacobs}, {Kaplan}, {Kim}, {Lenc}, {Line}, {Loeb}, {Lonsdale}, {Morales}, {Morgan}, {Neben}, {Thyagarajan}, {Oberoi}, {Offringa}, {Ord}, {Paul}, {Pober}, {Prabu}, {Riding}, {Udaya Shankar}, {Sethi}, {Srivani}, {Subrahmanyan}, {Sullivan}, {Tegmark}, {Webster}, {Williams}, {Williams}, {Wu}, \& {Wyithe}}]{trott16}
{Trott}, C.~M., {Pindor}, B., {Procopio}, P., {et~al.} 2016, \apj, 818, 139, \dodoi{10.3847/0004-637X/818/2/139}

\bibitem[{{Trott} {et~al.}(2019){Trott}, {Fu}, {Murray}, {Jordan}, {Line}, {Barry}, {Byrne}, {Hazelton}, {Hasegawa}, {Joseph}, {Kaneuji}, {Kubota}, {Li}, {Lynch}, {McKinley}, {Mitchell}, {Morales}, {Pindor}, {Pober}, {Rahimi}, {Takahashi}, {Tingay}, {Wayth}, {Webster}, {Wilensky}, {Wyithe}, {Yoshiura}, {Zheng}, \& {Walker}}]{kde2019}
{Trott}, C.~M., {Fu}, S.~C., {Murray}, S.~G., {et~al.} 2019, \mnras, 486, 5766, \dodoi{10.1093/mnras/stz1207}

\bibitem[{{Trott} {et~al.}(2020){Trott}, {Jordan}, {Midgley}, {Barry}, {Greig}, {Pindor}, {Cook}, {Sleap}, {Tingay}, {Ung}, {Hancock}, {Williams}, {Bowman}, {Byrne}, {Chokshi}, {Hazelton}, {Hasegawa}, {Jacobs}, {Joseph}, {Li}, {Line}, {Lynch}, {McKinley}, {Mitchell}, {Morales}, {Ouchi}, {Pober}, {Rahimi}, {Takahashi}, {Wayth}, {Webster}, {Wilensky}, {Wyithe}, {Yoshiura}, {Zhang}, \& {Zheng}}]{trott20}
{Trott}, C.~M., {Jordan}, C.~H., {Midgley}, S., {et~al.} 2020, \mnras, 493, 4711, \dodoi{10.1093/mnras/staa414}

\bibitem[{{van Haarlem} {et~al.}(2013){van Haarlem}, {Wise}, {Gunst}, {Heald}, {McKean}, {Hessels}, {de Bruyn}, {Nijboer}, {Swinbank}, {Fallows}, {Brentjens}, {Nelles}, {Beck}, {Falcke}, {Fender}, {H{\"o}randel}, {Koopmans}, {Mann}, {Miley}, {R{\"o}ttgering}, {Stappers}, \& {others}}]{vanhaarlem13}
{van Haarlem}, M.~P., {Wise}, M.~W., {Gunst}, A.~W., {et~al.} 2013, \aap, 556, A2, \dodoi{10.1051/0004-6361/201220873}

\bibitem[{{Wyithe} \& {Morales}(2007)}]{wyithe07}
{Wyithe}, J.~S.~B., \& {Morales}, M.~F. 2007, \mnras, 379, 1647, \dodoi{10.1111/j.1365-2966.2007.12048.x}

\end{thebibliography}
\bibliographystyle{aasjournal}



\end{document}